\documentclass[manuscript]{aastex}

\usepackage{graphicx}

\shorttitle{Searching for Trojan Asteroids in the HD 209458 System}
\shortauthors{Moldovan et al.}

\begin{document}

\title{Searching for Trojan Asteroids in the HD 209458 System:
Space-based MOST Photometry and Dynamical Modeling}

\author{Reka Moldovan\altaffilmark{1}, Jaymie M. Matthews, and Brett Gladman}
\affil{Department of Physics and Astronomy, The University of British Columbia, 6224 Agricultural Road., Vancouver, BC, V6T 1Z1, Canada}
\email{rmoldova@eos.ubc.ca}

\author{William F. Bottke}
\affil{Department of Space Studies, Southwest Research Institute, 1050 Walnut Street, Suite 400, Boulder, CO 80302, USA}

\and

\author{David Vokrouhlick\'y}
\affil{Institute of Astronomy, Charles University, Prague, V Hole\v{s}ovi\v{c}k\'ach 2, 180~00 Prague 8, Czech Republic}

\altaffiltext{1}{Currently at the Department of Earth and Ocean Sciences, The University of British Columbia, 6339 Stores Road., Vancouver, BC, V6T 1Z4, Canada}

\begin{abstract}

We have searched $Microvariability ~ and ~ Oscillations ~of ~STars$ (MOST) satellite photometry obtained in 2004, 2005, and 2007 of the solar-type star HD 209458 for Trojan asteroid swarms dynamically coupled with the system's transiting ``hot Jupiter'' HD 209458b.  Observations of the presence and nature of asteroids around other stars would provide unique constraints on migration models of exoplanetary systems. Our results set an upper limit on the optical depth of Trojans in the HD 209458 system that can be used to guide current and future searches of similar systems by upcoming missions. Using cross-correlation methods with artificial signals implanted in the data, we find that our detection limit corresponds to a relative Trojan transit depth of 1$\times10^{-4}$, equivalent to $\sim$1 lunar mass of asteroids, assuming power-law Trojan size distributions similar to Jupiter's Trojans in our solar system. We confirm with dynamical interpretations that some asteroids could have migrated inward with the planet to its current orbit at 0.045 AU, and that the Yarkovsky effect is ineffective at eliminating objects of $>$ 1 m in size. However, using numerical models of collisional evolution we find that, due to high relative speeds in this confined Trojan environment, collisions destroy the vast majority of the asteroids in $<$10 Myr. Our modeling indicates that the best candidates to search for exoTrojan swarms in 1:1 mean resonance orbits with ``hot Jupiters" are young systems (ages of about 1 Myr or less). Years of Kepler satellite monitoring of such a system could detect
an asteroid swarm with a predicted transit depth of $3 \times 10^{-7}$.

\end{abstract}

\keywords{minor planets, asteroids; planetary systems: formation; stars: individual(HD 209458)}
$Published~ by~ the~ Astrophysical~ Journal$, 714:315-323, 2010 June 10\\
\indent doi:10.1088/0004-637X/716/1/315

\section{Introduction}

\indent Currently, there are slightly over 440 known exoplanets (The Extrasolar Planets Encyclopaedia; exoplanet.edu). About 60 of these have orbital planes aligned with our line of sight so that the planets transit their parent stars causing small dips in brightness repeating at the planet's orbital period. Most of the known exoplanetary systems have gas giant planets in small, very short-period orbits, dubbed ``hot Jupiters" due to their large sizes and proximities to their parent stars.  Dynamical models \citep{laughcham2002, thommes2005, cressnel2006, lyra2009} predict that some of these hot Jupiters may be accompanied in their orbits by swarms of Trojan asteroids at the L4 and L5 Lagrangian points, like those in resonance with Jupiter in our own solar system \citep{stacon2007}. The detection and characterization of Trojan asteroids in an exoplanetary system would provide important constraints on the dynamical evolution of giant exoplanets.  

\indent Only a few years ago, the prospect of being sensitive to even a lunar mass of asteroids around another star would have seemed like science fiction.  The new detection limits made possible by space-based photometry missions like MOST, CoRoT, and Kepler mean for the first time that planetary scientists can seriously explore limits on the formation and evolution of possible exoTrojan swarms.  

\indent One way of detecting exoplanetary Trojans is through transit timing by combining radial velocity observations and photometric observations of a transiting planet. Through this approach a Trojan swarm can be detected via a time lag between the radial velocity null and the time of the central transit. Using this method, \citet{forgaudi2006} ruled out Trojan companions to HD 209458b of total mass greater than $\sim$ 13 M$_{\earth}$ at a 99.9\% confidence level. 

\indent In this paper, we present a more direct and more sensitive approach, by searching directly for the transit signal of Trojan swarms with precise spacebased photometry. Our target for this study was the HD 209458 system, which contains a hot Jupiter (designated HD 209458b). The planet was discovered in radial velocity measurements by \citet{mazeh2000} and the transits were first reported by \citet{charbonneau2000}.

\indent HD209458b has a circular orbit of semimajor axis $a = 0.045$ AU and period $P = 3.52474859 \pm 0.00000038$ days \citep{knutson2007}. The primary HD 209458 is a relatively-bright ($V = 7.65$) G0 star with $T_{eff} = 6000 \pm 50$ K, luminosity $L = 1.61$ L$_{\odot}$ \citep{mazeh2000}, mass $M = 1.10 \pm$ 0.07 M$_{\odot}$, and radius $R = 1.13 \pm$ 0.02 R$_{\odot}$ \citep{knutson2007}. Although quite uncertain, \citet{melo2006} give an estimate of 3 Gyr for the age of the system.  HD 209458 is a good starting point for the steps to compare the dynamical state of asteroids in an exoplanetary system with the current state of our solar system: the host star is Sun-like, and it is one of the brightest transiting systems available for study.

\indent It is widely accepted that hot Jupiters formed at orbital distances greater than their current locations and then migrated inward. We explore Trojan survival in such a system. To be visible today, the Trojan populations would have first had to survive a migration from about 5 AU down to about 0.05 AU. Once they arrived at this distance, both self-collisions and radiation effects (such as the Yarkovsky effect, which can cause small objects to undergo orbital changes) may be important for these hypothetical ``hot Trojans'' over times scales comparable to the system's age. We explore these effects through numerical simulations and discuss the results. 

\indent This paper is organized as follows. In Section 2 we describe how the photometric data were collected. In Section 3, we describe the data analysis, methods, and results. Section 4 is an estimate of the photometric cross section of the Trojan clouds in the HD 209458 system based on the solar system Trojan size distribution. Section 5 presents numerical calculations regarding dynamical and collisional evolution, as well as the Yarkovsky effect in this system. Finally, Section 6 provides conclusions based on these results. 

\section{Observations}

\indent The $MOST$ microsatellite \citep{walker2003, matthews2004} houses a 15 cm Rumak-Maksutov telescope feeding a CCD photometer through a single custom broadband filter covering roughly the visible part of the spectrum ($350 - 700$ nm). MOST was launched in 2003 June into an 820 km high circular Sun-synchronous polar orbit with a period of approximately 101.4 minutes. From this vantage point, it can monitor stars which lie in a roughly equatorial band (continuous viewing zone or CVZ) about 54$^\circ$ wide for up to 2 months without interruption. Photometry of very bright stars (visual magnitudes $V < 6$) is obtained in Fabry Imaging mode in which a Fabry microlens projects an extended image of the telescope pupil illuminated by the target starlight to achieve the highest precision \citep{matthews2004, guenther2008}. Fainter stars (down to about $V \sim 12$) can be observed in direct imaging mode, where defocused images of stars are monitored in subrasters on the CCD \citep{rowe2006}, similar to ground-based CCD photometry. 

\indent MOST monitored the transiting exoplanet system HD 209458 nearly continuously for 13.5 days in 2004 August (a trial run), 42.9 days in 2005 August to September, and 28.6 days in 2007 August to September. These data were used for several purposes: to measure the eclipse (and hence, albedo) of the ``hot Jupiter" \citep{rowe2006, rowe2008}; to perform transit timing of the known exoplanet to search for lower-mass planets in the system \citep{miller2008}; and to search for planets in the system approaching Earth-size through sensitive transit searches \citep{croll2007}.  

\indent HD 209458 was observed in direct imaging mode with exposures of 1.5 s sampled every 10 s (Rowe et al. 2008). The point-to-point precision for these observations ranges from as low as 3 mmag to as high as 20 mmag, depending on the level of stray light scattered into the instrument \citep{miller2008}. We rejected exposures with high cosmic-ray fluxes that occur when MOST passes through the South Atlantic Anomaly (SAA), as well as data with background illumination values greater than 3000 detector counts due to scattered earthshine modulated at the satellite's orbital period.

\indent The raw and reduced data are available in the MOST Public Data Archive (www.astro.ubc.ca/MOST/data/data.html). The time series photometry is presented in Figure 1, and the data phased according to the orbital period ($P \sim 3.5248$ d) of the exoplanet HD 209458b is shown in Figure 2. \

\section{Searching for Trojans}

\indent 
Since Trojan asteroids are expected to concentrate around the L4 and L5 Lagrangian points of the HD 209458 star-exoplanet system, they will share an orbit with the exoplanet. The L4 and L5 points themselves transit the star 1/6th of an orbital period behind and ahead of the exoplanet, with the clouds occupying a confined range of angles surrounding these points. It is expected that these clouds will occupy a fairly wide area (at least $\pm 30^{\circ}$ about the L4/L5 point, as in the case of Jupiter's Trojan clouds) and will therefore have a wide signature in the light curve, each occupying $\simeq0.2$ of the orbital phase. The multitude of small asteroids would block a small fraction of the stellar light during the period in which the cloud occupies the line of sight. Thus in the simplest case scenario this would look like a box-function dip centered on the L4/L5 point extending about 30$^{\circ}$ in each direction. More complicated models (covering a larger angular range, or having a Maxwellian profile, for example) are possible, but are not currently justified by the non-detection from the present data set.

\indent We first examined in more detail the particular phases in the phase diagram shown in Figure 2 where Trojan swarm transits are expected. The exoplanet transits and other obvious outliers were excised from the light curve before our analysis. As a first step, we divided the phased data into six non-overlapping bins, ensuring that the centers of two of those bins would correspond to the expected center phases of the two possible Trojan swarms. We calculated the mean relative flux value of each of the six bins. This accentuates any possible Trojan signals, as well as having the highest practical signal-to-noise ratio. This exercise, however, did not reveal a significant dip in the binned phase diagram at either the L4 or L5 points. 

\indent We then generated running means of the phased data with the width of the phase bins and the bin shift interval as input parameters. A representative result is shown in Figure 3, where the phase bin width is 0.05 cycle and the sampling interval is 0.001 cycle.  Since in this case, $96\%$ of the data in adjacent bins are the same, the resulting means are highly correlated. Figure 3 shows that there are local minima near both Trojan points; however, they are not unique, nor is their extent wide enough to signal Trojan transits.

\indent We tested our sensitivity to Trojan transits in the data by inserting synthetic transits of known duration and depth.  A Trojan swarm may have an irregular spatial distribution and hence a more complex transit signature, but for simplicity, we inserted transits of uniform depth $\pm$ 30$^\circ$ in size, and decided if we could detect these transits in a diagram like Figure 3. These tests indicated that our detection limit for Trojan transit depths 
corresponds to a relative magnitude drop of order $10^{-4}$. We then tackled our detection limits in a more quantitative way via cross-correlation.  
The time series data were split into a first half and second half of our available time series. The two halves were phased and binned in non-overlapping bins to ensure that there was only one flux value for each phase value. The two sets were also filtered with a low-pass digital Butterworth filter to eliminate any excess ringing in the data that could potentially hide Trojan signals. Then the two data halves were cross-correlated with each other. Figure 4(a) shows the two filtered data halves, while Figure 4(b) shows the cross-correlation.  The mean parametric confidence limits were calculated using the MATLAB code ``xcorrc" \citep{saar2004}. Confidence limits are obtained by performing the cross-correlation in the Fourier domain. 
The code takes the discrete Fourier transform of both data sets, complex conjugates one of them, and then to find the confidence intervals, it introduces chance by randomly replacing the phase values of this second data set. This is then multiplied by the first data set. This procedure is repeated $n$ times, thereby creating a distribution of cross-correlations for each lag value.  For example, to set the $95\%$ confidence interval, xcorrc finds at what level for each lag $95\%$ of the correlations are below that level (i.e., the level beyond which only $5\%$ of the peaks are greater strictly by chance). 

\indent Any real signal at constant phase with the planet's orbit should have a strong self-correlation at zero relative cross-correlation lag. In addition, if the phases around both L4 and L5 were to generate dimming, there should also be (weaker) cross-correlation centered close to phase lags of $+0.33$ and $-0.33$ (we will illustrate this with artificial signals below). No feature is present in Figure 4 at a correlation lag of zero, nor are broad features centered on $+0.33$ and $-0.33$ visible. The weak signal at a phase lag of $-0.1$ is not significantly stronger than what we expect due to random chance and cannot
be a Trojan signal in any case.

\indent We repeated the cross-correlation analysis with artificial transits of known depth inserted into the MOST photometry. The artificial transits were again simple box functions of widths $\pm30^{\circ}$ centered on the L4 and L5 points. By varying the depth of the artificial transits and calculating the $92\%$, $95\%$, and $99\%$ confidence intervals, we estimate the effective Trojan detection limit of our data. 

\indent To set a number for our detection limit, we calculated what percentage of the correlation peaks in the original data and each synthetic transit scenario were above the 99$\%$ confidence line. Since for each cross-correlation only $\sim1\%$ of the coefficients should be above the 99$\%$ strictly by chance; if this percentage is significantly higher, we can conclude that we are detecting genuine signal in the data. In addition, for it to be a Trojan signal we also require that a strong signal exists near zero phase, and possibly that there are peaks confined around the lags where the Trojans are expected. (The Trojans are expected at lags of approximately $+1/3$ and $-1/3$, since the L4 and L5 points are 1/3 in phase from each other; these peaks occur when lagged data points at L4 line up with points at L5, and vice versa). 

\indent Figure 5 shows a few sample cross-correlation results for three artificial Trojan transit tests. For an artificial transit of depth $8 \times 10^{-4}$ (Figures 5(a) and (b)), we recover very strong correlation at zero phase lag, and marginally-significant signals peaking at the expected $\pm$0.33 lags. If only one Trojan point were populated, the side lobes would disappear but the strong zero-lag signal would remain. An acceptable signal from artificial Trojans with dips of $2 \times 10^{-4}$ is shown in Figures 5(c) and (d). The zero-phase correlation is strong but, as expected, the $\pm0.33$ peaks have become less pronounced for shallower depths of transit (in fact, one of them is only barely significant). For a transit depth of $5 \times 10^{-5}$ (Figures 5(e) and (f)), only very weak correlations remain, due to the much reduced signal to noise, and we judge this a non-detection since the zero-phase signal is no stronger than the $-0.33$ feature. From this analysis, we conclude that our detection limit is $\simeq 10^{-4}$, the same as from our eyeball analysis of the data.

\section{Trojan Size Distribution $-$ Solar System Model}

\indent To estimate the flux reduction due to possible Trojan clouds in the HD 209458 system, the known properties of Jupiter's Trojan swarms in our solar system were applied to the HD 209458 system. In this way, we estimate the expected transit depth if the Trojans in the HD 209458 system had the same size distribution. 

\indent We first calculated the cross-sectional area of the Trojan clouds given this size distribution, as well as the mass of the swarms, and from there the transit depth was established. 

\vspace{0.25in}

\subsection{Cross-sectional Area of a Trojan Cloud}

\vspace{0.25in}

\indent We estimated the cross-sectional area of a single Trojan cloud by integrating the relation
\begin{equation}
da = \pi r^2 n dr,
\end{equation}
 \noindent where the differential size distribution $ndr$ measures with one power law for radii in the range 2.2 km $\leq r \leq 42$ km and another for $r \geq 42$ km \citep{jewitt2000}.
\noindent For radii smaller than $r \leq 2.2$ km, we calculated a differential size distribution by matching the Jewitt et al. result with that of \citet{yoshida2005} (see Figure 6). For $r \geq 42$ km, \citet{jewitt2000} give
\begin{equation}
n_{r>42}(r)dr = 3.5 \times 10^9 \left({\frac{1{\rm km}}{r}}\right)^{5.5 \pm 0.9}dr.
\end{equation}
\noindent In the range 2.2 km $\leq r \leq 42$ km, Jewitt et al. give
\begin{equation}
n_{42>r>2.2}(r)dr = 1.5\times 10^6  \left({\frac{1{\rm km}}{r}}\right)^{3.0 \pm 0.3} dr,
\end{equation}
\noindent which matches at the 42 km break. \citet{yoshida2005} find a cumulative power-law slope of 1.3 for Trojans with radii between 1 km $\leq r \leq 2.5$ km. Here we force a match at $r = 2.2$ km, adopting $n_{42~>~r~>~2.2}dr = n_{2.2~>~r}dr = 1.41 \times 10^5 dr$, from which we calculate the third power-law $n_{r~>~2.2}dr$ as
\begin{equation}
n_{2.2>r}(r)dr = c \cdot \left({\frac{1{\rm km}}{r}}\right)^{2.3 \pm 0.1}dr, 
\end{equation}
\noindent where $c = 8.65 \times 10^5$ km$^{-1}$. To find the total cross-sectional area of the Trojans, we substitute these distributions into Equation (1) and integrate to find the total cross-sectional area $a_{trojans} = a_{r~>~42} + a_{42~>~r~>~2.2} + a_{2.2~>~r}$. Using the best estimates for the slopes, the cross sections from the three distributions are
\begin{equation}
a_{r>42} = 3.5 \times 10^9 \cdot \pi \int^{\infty}_{42} {\left(\frac{1{\rm km}}{r}\right)^{5.5} r^2} dr \approx 1\times 10^6~ {\rm km}^2,  
\end{equation}
\begin{equation}
a_{42>r>2.2} = 1.5 \times 10^6 \cdot \pi \int^{42}_{2.2} {\left(\frac{1{\rm km}}{r}\right)^{3} r^2} dr \approx 1.4\times 10^7~ {\rm km}^2,
\end{equation}
and
\begin{equation}
a_{2.2>r} = 8.65 \times 10^5 \cdot \pi \int^{2.2}_0 {\left(\frac{1{\rm km}}{r}\right)^{2.3} r^2} dr \approx 7\times 10^6~ {\rm km}^2.
\end{equation} 

\indent While the asteroids smaller than 2.2 km do not contribute significantly to the mass of the Trojan clouds (see below), they do contribute significantly to the cross-sectional area and therefore the depth of transit.  Our estimate of the total cross-sectional area of a Trojan cloud is $a_{trojans} \approx 2\times 10^7$ km$^2$. Although it is conceivable that our Trojan cloud's optical depth could be governed by a large population of $r\ll1$~km particles which are currently unobservable, given the lack of evidence for this hypothesis we shall use $2\times10^7$~km$^{2}$ as our estimate of the total Trojan cross-section. Beyond this concern, the dominant source of uncertainty is the power-law indices in the size distribution, producing a variation of a factor of a few in the total cross section if we instead adopted a size distribution similar to \citet{ferni2009}.

\vspace{0.25in}

\subsection{Mass of Trojan Cloud: Our Solar System's}

\vspace{0.25in}

\indent We calculate the total mass of Trojans by integrating the three differential size distributions and assuming a mean asteroid density of $\bar {\rho} = 2000$ kg m$^{-3}$. In this way, \citet{jewitt2000} estimated the total mass of solar system Trojans with $r > 2.2$ km as $M_T \approx 5\times10^{20}$ kg. We extend this result by including the third differential size distribution $n_{2.2~>~r}dr$. 
The total mass of Trojans is then
\begin{equation}
 M_T = \int^{2.2}_0{\frac{4}{3} \pi \bar{\rho} r^3 n_{2.2>r}}dr
  + \int^{42}_{2.2}{\frac {4}{3} \pi \bar{\rho} r^3 n_{42>r>2.2}}dr\nonumber \\
 + \int^{\infty}_{42}{\frac{4}{3} \pi \bar{\rho} r^3 n_{r>42}}dr. 
\end{equation}

\noindent This yields $M_T\sim5.9\times10^{20}$~kg, containing (as expected) negligible additional mass in the distribution tail with $r~<~2.2$ km. Expressed in lunar masses (M$_{Moon}\sim7.36\times10^{22}$~kg), this is about 0.008 M$_{Moon}$. This mass estimate should be halved if the measured density of $\rho\sim$1000~kg m$^{-3}$ for the Trojan Patroclus is common \citep{marchis2006}.

\vspace{0.25in}

\subsection{Depth of Transit}

\vspace{0.25in}

\indent To estimate the drop in measured flux of the star due to a population of Trojans as described above, we assume that Trojans with that size 
distribution are orbiting the star HD 209458a surrounding the L4 and L5 points of HD 209458b. Given our null detection, we believe only order-of-magnitude estimates are warranted at this time. We take the star's radius to be $R_{star} = 1.13R_{Sun} = 7.85 \times 10^5$ km, and assume Trojans to be evenly distributed around the L4 and L5 points. If all the L4 Trojans, or all the L5 Trojans (whose populations we take to be equal and thus half the total), were in front of the star at the same time, the drop in light output would be 

\begin{displaymath}
\frac{\Delta I}{I} = \frac{a_{trojans}/2}{a_{star}} \sim 6\times10^{-6},
\end{displaymath} 

\noindent where we are neglecting the effects of limb darkening. However, we need to account for the fact that the longitudinal extent of the Trojan cloud at either of the Lagrangian points will most likely be larger than the angular size of the star. Very roughly, the longitudinal extent of Jupiter's Trojans around one Lagrangian point is about $\pm$ $30^{\circ}$. The projected stellar diameter, expressed in terms of the exoplanet's orbit, is the fraction of that orbit covered by the planet's transit, which is 0.035 in phase ($13^{\circ}$). This means that a Trojan cloud like that of Jupiter in the HD 209458 system has only about a fifth of its Trojans in front of the star during transit. Therefore, we divided our drop in magnitude by a factor of 5 giving $\frac{\Delta I}{I} \sim 1.2\times10^{-6}$. 

\indent Another issue that may affect the star's brightness is that Jupiter's Trojans have a large dispersion above and below the ecliptic plane. From \citet{yoshida2005}, the average Trojan inclination is $10^{\circ}$, and we take the putative exoplanet Trojan system to be $\sim30^{\circ}$ (where Section~\ref{sec:mig} motivates the factor of 3 increase). Knowing that the Trojan semi-major axis is $a = 0.045$ AU and that the star's radius is $R_{star} = 0.0052$ AU, the  ``height" of the Trojan cloud is $\sim$5 stellar diameters above and below the orbital plane. Thus we divide our $\frac{\Delta I}{I}$ by 5, although this effect will not occur if the HD 209458b Trojan orbits have inclinations significantly $<30^{\circ}$. Our final estimate for the drop in the star's flux due to the transit of a Jovian-type Trojan cloud is $\frac{\Delta I}{I} \sim 3\times10^{-7}$. To detect such a transit requires a photometric precision of order $3\times10^{-7}$ assuming the Trojans obey the solar system size and angular distributions. 

\subsection{Exotrojan Mass}

\indent Based on the above calculations, we estimate the mass of the hypothetical Trojan cloud in the HD 209458 system that would be required for MOST to have a detection. Since our detection limit for a Trojan transit is a drop in the light output of 1$\times10^{-4}$, we would not be able to detect the $3\times10^{-7}$ drop caused by a Trojan population of the size of Jupiter's Trojan clouds. However, for
\begin{displaymath}
\left. \frac{\Delta I}{I} \right|_{limit} = 1\times10^{-4} = \frac{a_{trojans}}{a_{star}},
\end{displaymath}
so we find that the Trojan cross-sectional area is $a_{trojan} \sim 2\times10^8$km$^2$, which is a factor of $\sim$100 times the optical depth of Jupiter's Trojan cloud. Therefore MOST could detect a Trojan cloud in front of HD 209458a if the number of asteroids in each size bin (assuming the size distribution of Jupiter's Trojans) were increased by a factor of 100. This would increase the mass of the entire cloud by this factor, thus bringing the minimum exotrojan mass that we could detect to $\sim$1 lunar mass.

\section{Dynamical effects}

\indent In this section we discuss the relationship between the upper limit on the HD~209458b Trojan population given by our non-detection and what one might expect to exist in the exoplanet system. In order to have some concreteness (in what is an otherwise large and unconstrained parameter space), we will mostly concentrate on the scenario in which HD~209458b forms at $\sim$5~AU from its star along with a Trojan population equivalent to some multiple of Jupiter's current population, after which it migrates to its current position and then the Trojans and planet remain at the current stellar distance. 

\indent The efficient mechanisms for planet migration, in which the planet couples to the gas disk, would require that the planet migrated to its current position within the first $\sim$3--10 Myr of the system's lifetime, before the circumplanetary gas was dissipated. Since HD~209458 is $\sim$3~Gyr old, the time that the system has been at its current orbital distance is 2--3 orders of magnitude longer than the migration phase.

\subsection{Migration phase}
\label{sec:mig}

\indent As the planet migrated in toward the star the 1:1 resonant Trojans also spiraled inward, remaining trapped in the resonance although their libration amplitudes grew. (The libration amplitude $A$ measures the total amplitude of the angular variation of a given particle away from the Trojan point in the reference frame co-rotating with the planet). We have conducted straightforward numerical simulations to confirm the result of \citet{flamham2000} that the libration amplitude slowly grows as the planet migrates inward according to:
\begin{equation}
A_f = \left(\frac{a_f}{a_i}\right)^{-1/4} A_i,
\end{equation}

\noindent
where $a$ is the semimajor axis of the planet's orbit and subscripts $i$ and $f$ indicate the initial and final values of the variables (before and after migration). \citet{flamham2000} show that the eccentricity and inclination of the Trojan orbits grow by the same factor. Assuming HD 209458b began about 5~AU from the star, its current semimajor axis results in Trojan libration amplitudes growing by a factor of $\sim 3.3$ while migrating, which is confirmed in our numerical simulations. This results in a large fraction of the Trojan phase space having libration amplitudes that grow beyond the maximum possible stable value (of about $130^{\circ}$) and leave the resonance, after which they will interact with the planet and be accreted or ejected. Using a uniformly-filled initial Trojan phase space results in $\sim 10\%$ of the Trojans (those with initial libration amplitudes $<35^{\circ}$) surviving migration, where the resultant stable co-orbitals have libration amplitudes that have grown to fill the stable libration region. Given what follows below, improved precision on these estimates is not currently warranted.

\subsection{Collisional Evolution}

\indent A small-body population, like our asteroid belt, will have its population and size distribution evolve collisionally if the spatial density is sufficiently high and relative speeds are large enough that collisions are frequent and cause net erosion.  
In the case of a hypothetical HD 209458b cloud, collisions could be occurring  in the pre-migration phase, during the planetary migration, and then in the post-migration period. Since the final stage is both much longer and has the highest relative speeds and spatial density (since the Trojans occupy much less volume when they surround the Trojan point at 0.045~AU than at $\sim$5~AU), we will neglect the collisional evolution except during the post-migration
phase. This collisional evolution will result in collisional fragmentation of Trojans, which will move mass from large objects in a size distribution down to smaller diameters; when collisions produce particles small enough that radiation-pressure effects become important, they are quickly eliminated from the system. Collisions will thus grind down the total mass of the system on some time scale. In our solar system the internal collisional cross section is low enough that the main asteroid belt's mass has not significantly been reduced by grinding over the last 4 billion years \citep{bottke2005}. 

\indent We posed the questions: If a Trojan cloud survives migration down to 0.045~AU, how would it evolve? If collisions are important, can we postulate an increased initial mass in order to obtain enough Trojans today to have an observable signal? Recall that Jupiter's Trojan cloud has a mass $\sim0.01$ lunar masses, so we will use order-of-magnitude multiples of this for hypothetical populations. Trojan clouds with initial orbital eccentricities and inclinations similar to our solar system Trojans were used, which determines the collision speeds to be $\sim v_{kep}\sqrt{e^2 + i^2} \sim 65$ km s$^{-1}$, where $v_{kep}$ is mean orbital speed of the Trojans. If migration results in even higher typical $e$'s and $i$'s, then these speeds would grow by a factor of 3--4. This enormous mutual velocity means that collisions are very destructive when compared to those occurring in our asteroid belt, where mean speeds are only about 5~km s$^{-1}$ \citep{bottke1994}. Although this mutual speed could be decreased if the mutual $e$'s and $i$'s were dropped, the mutual collisional cross section increases under this change because the volume occupied by the cloud decreases.

\indent The Trojan swarm was evolved collisionally using an algorithm very similar to that described in \citet{morbidelli2009}, which keeps track of the evolving size distribution of the collisional swarm. Figure 7 shows the evolution of the size distribution between 0.1 and 1000 km at various times in the evolution for an initial 1 lunar-mass population, with an initial differential power-law size index of $-3$. The population is rapidly decimated by mutual collisions, which by 10 Myr leaves almost all the mass in a single 35 km (radius) object while reducing the number of 1 km objects (where most of the cross section is) by more than 8 orders of magnitude. Thus, even if an initial Trojan swarm of 1000 times that of Jupiter started at 5~AU and then arrived at 0.045~AU in a migration that reduced it by an order of magnitude, in only 10 Myr the light-blocking cross section is reduced by more than a factor of a million, making detection today many orders of magnitude below current or projected flux limits for HD 209458b. A few million years into the simulations, the total Trojan cross-sectional area is $\sim10^{-8}$ that of the star, and continues to decrease.

\indent Neither increasing nor decreasing the mass of the cloud that arrives at 0.045~AU helps. Additional simulations with 1, 2, or 3 orders of magnitude more or less initial mass arrive at about the same final state after 10 Myr of collisions, due to the simple fact that the collision rate is proportional to the number of objects. In fact, the total amount of mass after 10 Myr is comparable to the 0.01 Myr state (see Figure 8) since the system grinds itself down to the point where the collisions `turn off' because the remaining objects have so little mutual cross section.

\indent Experiments with $-4$ power-law indices gave similar results. Our conclusion is that, unless the collisional modeling is wrong by many orders of magnitude, Trojan clouds with detectably large cross-sectional area will not survive for even a million years after the planet migrates to hot Jupiter distances. Hiding the mass in a smaller number of large bodies would prevent collisions from destroying the mass, but these bodies have such small cross-section/mass ratios that such a population would have negligible optical depths and cannot be detected via light-curve technology in the foreseeable future.\\

\subsection{Emptying the Trojan points}

\indent The Trojan population which initially arrives with the planet at 0.045~AU at the end of the migration phase is initially modified by collisions. The collisional cascade fragments larger bodies and populates the smaller-diameter bins, whose bodies in turn are destroyed by collisions among smaller particles. In our asteroid belt this process stops when particles are ground down to sub-millimeter size at which point radiation forces eliminate them, or when bodies of 0.01--1 km scale have their semimajor axes slightly modified by Yarkovsky drift which moves them into a resonance at which point they
can rapidly leave the belt (reviewed in \citet{bottke2006}). The migration rates induced by these processes and migration direction (toward or away from the star) depends on the particle's size, spin rate, obliquity, and the thermal properties of the surface.

\indent In the case of exoplanet Trojans, these forces do not cause a gradual monotonic semimajor axis drift (which would be $\sim \pm 1$~mm s$^{-1}$ for a 1 m diameter rocky object). A straightforward analysis of the circular restricted three-body problem shows that the additional constant acceleration, which would produce a slow semimajor axis change outside the resonance, instead causes a tiny change ($<1$ part in $10^4$) in the positions of the Lagrange points. This effect is analogous to how a small damping force added to a driven harmonic oscillator only produces a phase shift.

\indent This shift in the location of the libration centers is not in itself a source of instability, and thus the radiation damping would not cause the particles to leave the Trojan points. However, because the libration center moves if the object suffers a collision which changes its spin rate and direction, the libration amplitude will random walk toward larger values and result in the eventual destabilization of the Trojan; we estimate this time scale to be $\sim 1$~Gyr for metre-scale Trojans of HD 209458b. We showed above that the rapid collisional grinding efficiently transfers mass from larger objects ($>1$~km) down to the regime where radiation effects can then lead them to be destabilized and pulled down to the star. In particular, once free of the resonance, metre-scale objects at these distances will spiral into the star from Poynting-Robertson drag in only 2 Myr, with the time scale proportional to the object size \citep{gladcof2009}. As a result of these processes, there may be a period during the system evolution where a large amount of mass has moved into the diameter region just above where radiation-pressure effects are efficient at eliminating small particles. This will be the size regime with the best optical depth to mass ratio, and it is possible that a nearly-opaque cloud in one or both Trojan points could be temporarily produced. Such a cloud would have a very strong photometric signature (essentially blocking the stellar light for the portion of the disk that is eclipsed, for the portion of rotational phase that it is in front of the star). Our collisional simulations above indicate that this would be a brief (certainly $<$ 1~Myr) phase that would most likely occur in a very young system during the grinding phase before the mass is eliminated. (In fact it is possible that it might occur during the migration phase itself). This could also occur in older systems if a large body manages to survive in a Trojan point but is then broken up, beginning a similar rapid collisional cascade at that time.\\

\section{Conclusions}

\indent In this paper, photometric data on HD 209458 from the MOST space telescope were analyzed with the intent of finding Trojan asteroids swarms in 1:1 mean motion resonance with the known planet HD 209458b. Using cross-correlation techniques on data with synthetic Trojan cloud transits, we were able to set an upper limit on the Trojan transit depth of 1$\times10^{-4}$, which corresponds to an upper limit in asteroid swarm mass to $\sim$ 1 lunar mass. 

We then assessed the dynamical effects on such a Trojan cloud in the HD 209458 system. The main findings are as follows.

\begin{itemize}

\item During the migration of the Trojan swarm (as they migrate with the planet from $\sim$ 5 AU to 0.045 AU), the libration amplitudes grow by a factor of $\sim$ 3.3 leaving $\sim 10\%$ of the Trojans surviving migration.

\item The collisional evolution of the cloud grinds it down to below 100 m size objects in $\ll$10 Myr and reduces its surface area 10,000 fold. Neither increasing nor decreasing the initial mass of the cloud changes the results significantly. After only a few Myr the total cross section remaining in the Trojan swarm (in bodies larger than tens of meters) has dropped to $<10^{-8}$ that of the star.  Based on the initial photometric performance of Kepler long cadence
data \citep{jenkins2010}, for a solar-type
star of V $\sim$ 10 with Trojan asteroid swarms in a 3.5 day orbit,
 the light curve would reach a sensitivity
to transit depth of $3 \times 10^{-7}$.  In a young system, this
would be sufficient to detect the exoTrojan swarms based on our model
predictions.

\item Radiation forces do not cause a semimajor axis drift of the leftover small objects, only a shift in the location of their libration centers. Further collisions between these small objects could then cause a random walk in their libration amplitudes, eventually destabilizing the Trojans 
and eliminating them on a time scale of 2 Myr. 

\item A nearly opaque cloud might result from the collisionally ground asteroid population just before the radiation forces take effect. Such a cloud would have a very strong photometric signature for a brief time period ($<$ 1 Myr) due to the dense population of small particles. This would occur during the collisional phase of the cloud evolution, and since this is a fast process, this could only be observed in a very young system ($\ll$100 Myr).

\end{itemize}

\indent Thus, the best chance of detecting Trojan asteroid swarms in another solar system in the foreseeable future would be in very young systems, where either the collisional depletion has not yet reduced the surface area significantly, or where the cloud turned opaque for a short time due to the high density of centimeter-sized objects before elimination by Poynting-Robertson drag. Detection in an older system would require a recent breakup of a large body that generated a collisional cascade of smaller objects. 

\indent Our simulations do rely on the assumption of simple migration of the gas giant planet.  Other processes, involving planet-planet interactions and later tidal circularization of the exoplanet's small orbit, could lead to different scenarios of exoTrojan migration and evolution.  However, we note that most of the observations of the Rossiter-McLaughlin effect in exoplanet systems, including HD 209458 \citep{winn2005}, have pointed to only small levels of spin-orbit misalignment in systems with close-in giant planets \citep{winn2007, narita2009}.  There is no reason to strongly suspect planet-planet interactions in the case of HD 209458 and other hot Jupiter systems with small, nearly circular orbits. 

\indent What about exoplanets in orbits with much larger semi-major axes?  For exo-Jupiters with relatively large semi-major axes, the prospects of detecting transits in the existing sample are limited. The Kepler mission will expand the sample and should change that situation in the coming years. Many of the known exoplanets with larger semi-major axes also have high eccentricities. In these systems, there could have been dynamical interactions with other planets earlier in the histories of the systems.  We argue that these are not systems where it would be fruitful to search for Trojans in resonant orbits, and for that reason, we did not explore this broader range of parameter space in the numerical simulations presented in this paper.

\indent Our analysis of ultra-precise MOST photometry provides the most sensitive upper limit yet published on the amount of Trojan asteroidal material in an exoplanetary system. This upper limit inspired us to perform numerical simulations to show what Trojan optical depth might be present.  By estimating the expected optical depth of an exoplanetary asteroid cloud, we have set meaningful limits on future photometric searches for exoTrojan swarms by missions such as Kepler, and focus attention in particular on very young systems.

\acknowledgments

R.M., J.M.M., and B.G. acknowledge the support of NSERC.  We thank the referee for insightful comments which improved the paper.

\pagebreak

\begin{figure}[ht]
\begin{center}
\includegraphics[scale=0.60, angle=0]{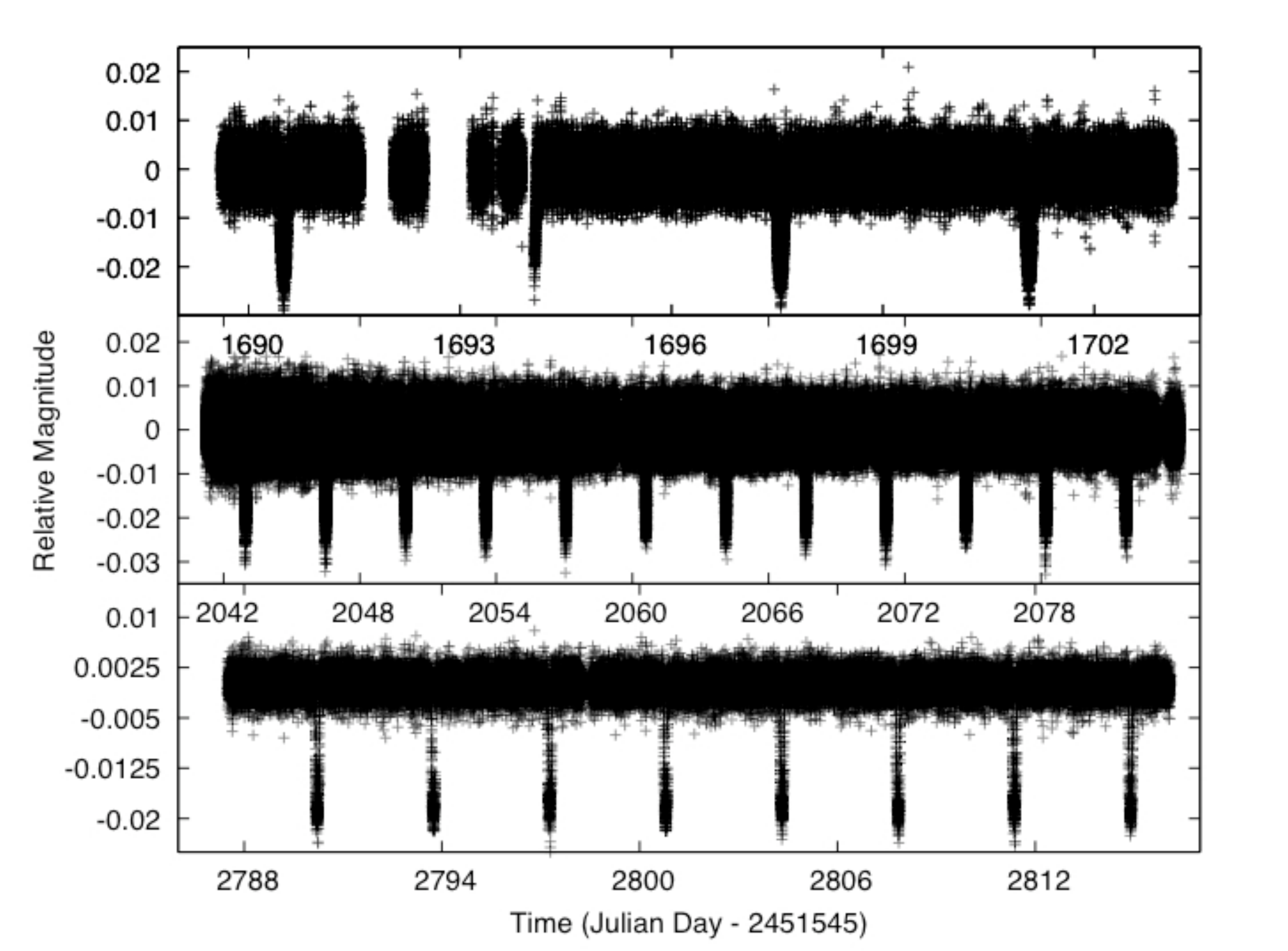}
\caption{ MOST photometry of HD 209458 (from observations in 2004, 2005, and 2007, respectively) presented as flux variations relative to the mean. }
\end{center}
\end{figure}

\begin{figure}[ht]
\begin{center}
\includegraphics[scale=0.25, angle=0]{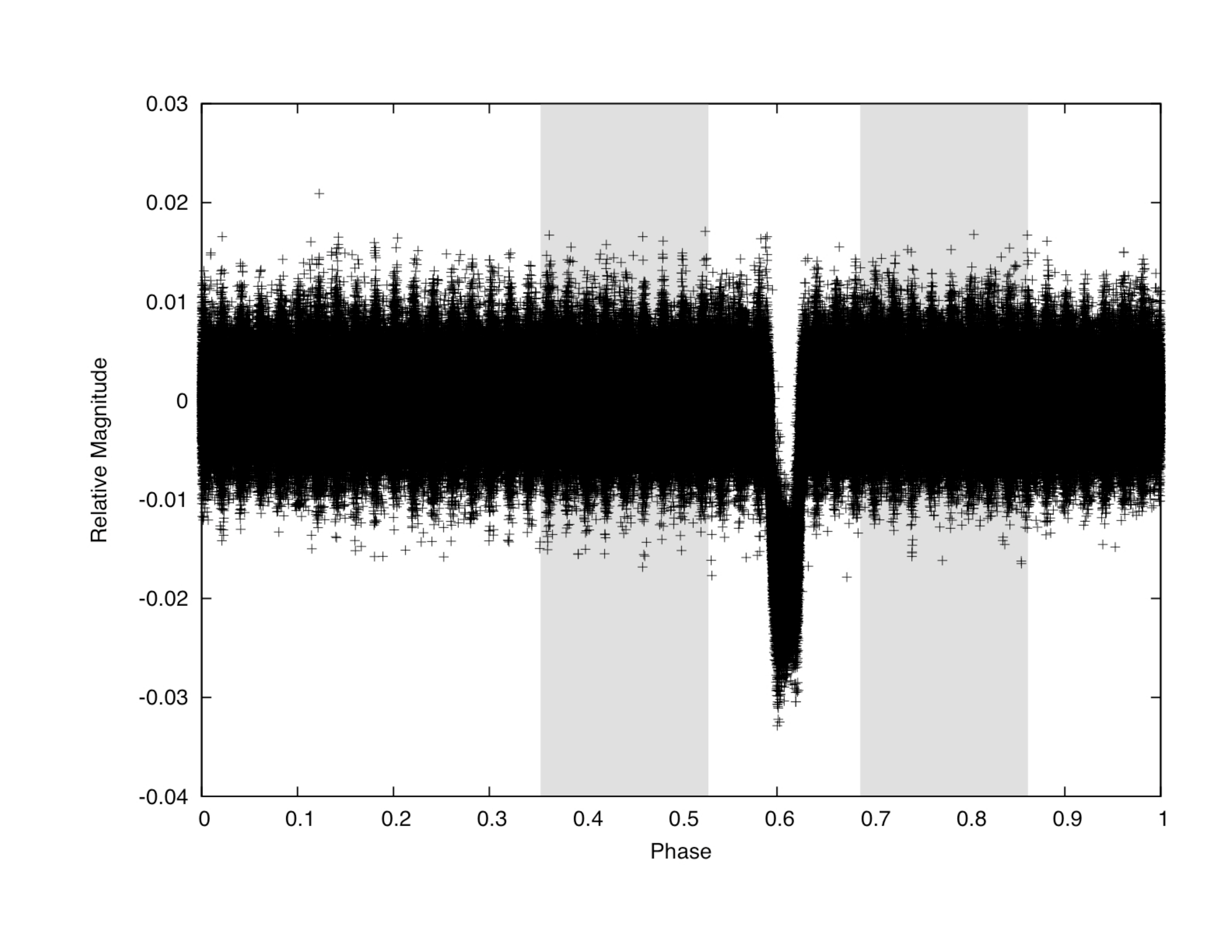}
\caption{Phase diagram of MOST photometry of HD 209458. Phase is judged relative to the start of the time series, folding at the known planetary orbital period. The planetary transit is obvious. The shaded areas represent sections of the phase diagram where dips in the light curve due to Trojan transits would be expected.}
\end{center}
\end{figure}

\begin{figure}[ht]
\begin{center}
\includegraphics[scale=0.8, angle=0]{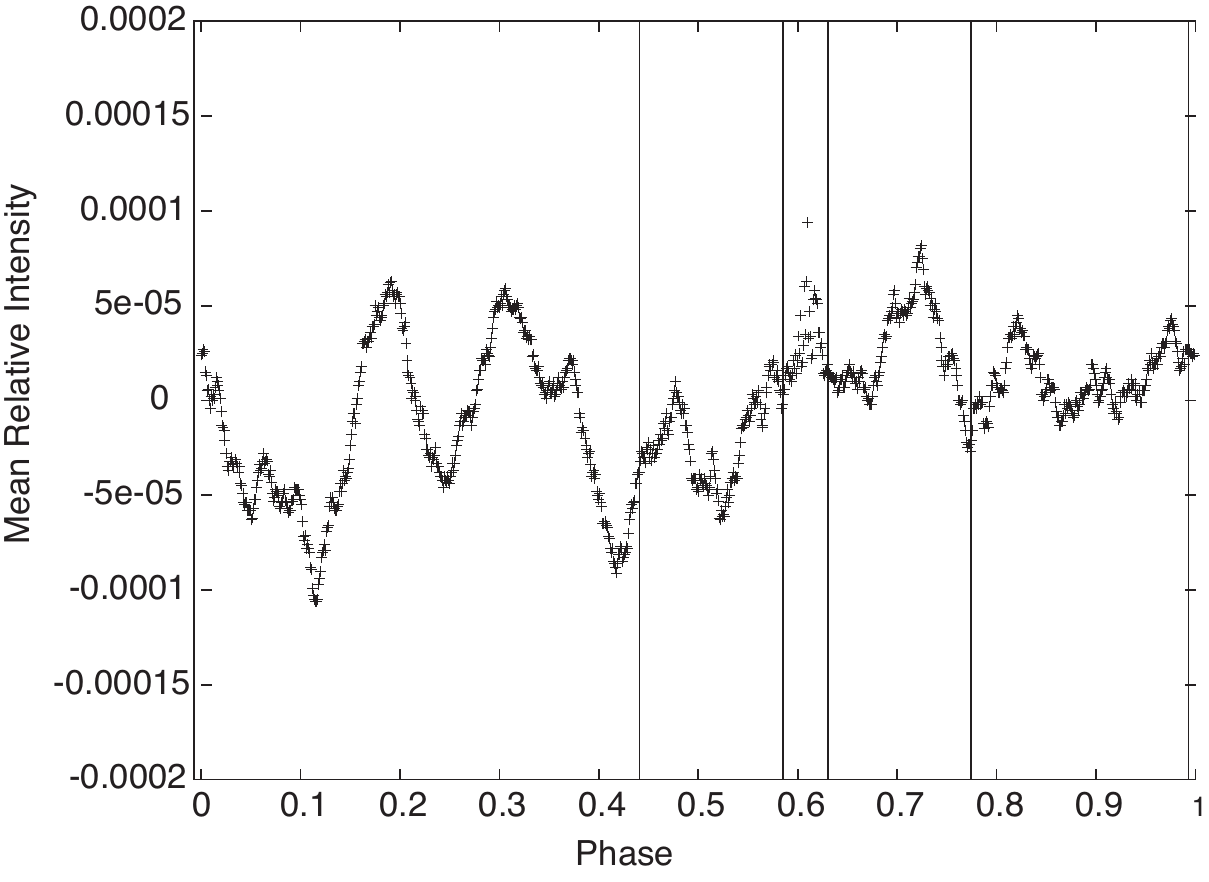}
\caption{Phase diagram of MOST photometry of HD 209458, binned in overlapping bins of width 0.05 of the orbital cycle, where the bin centers are spaced by only 0.001 cycle. Because of the use of a running average, the noise properties of the signal are masked here; however, if there was a broad ($\simeq 0.2$in phase) signal due to a transit, it would be apparent in the figure. Orbital phases corresponding to the exoplanet transit (near 0.6) and the transit of the precise L4 and L5 points 
are indicated by vertical lines.}
\end{center}
\end{figure}

\begin{figure}[ht]
\begin{center}
\includegraphics[scale=1, angle=0]{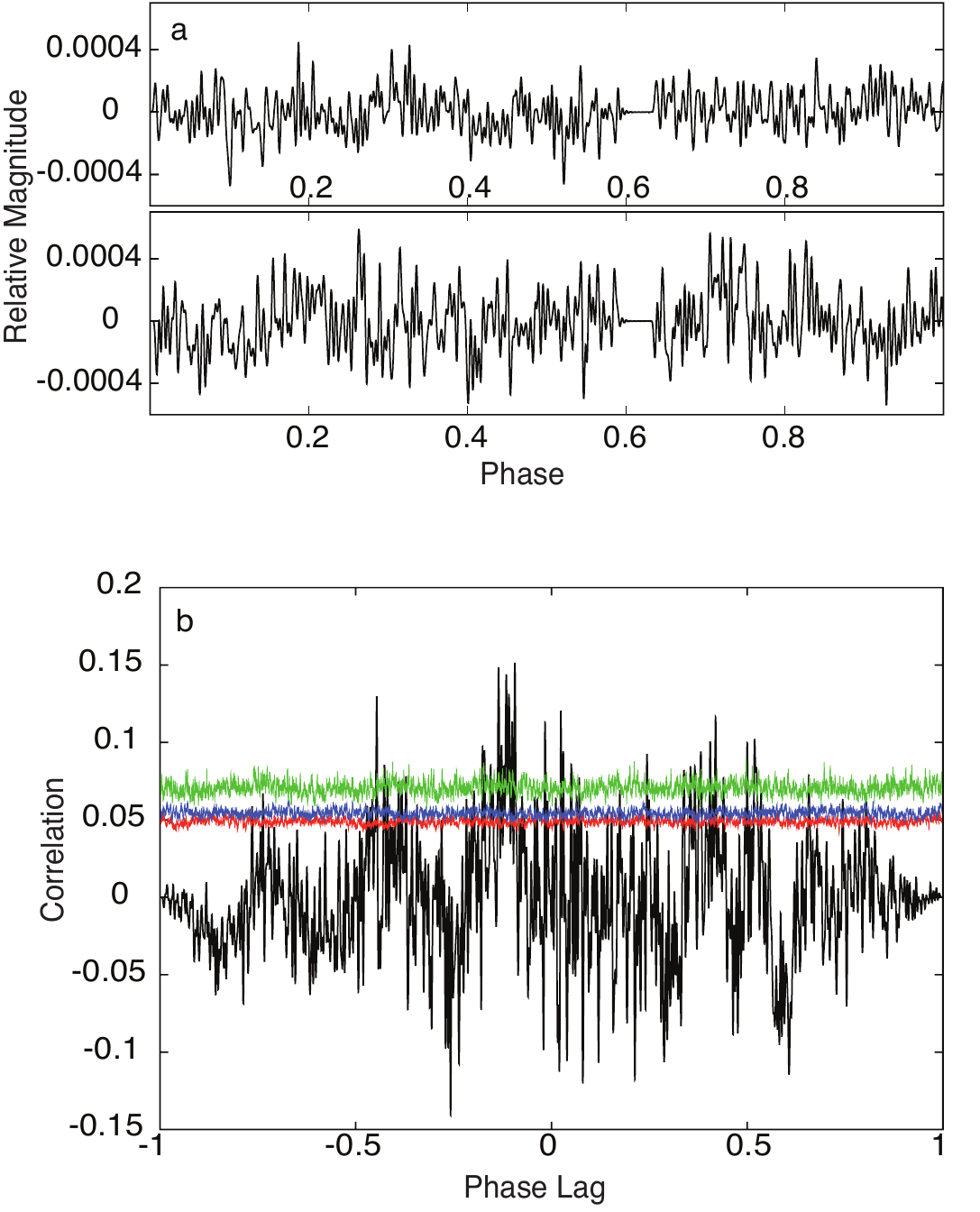}
\caption{(a) The two data halves before cross-correlation filtered using a normalized cutoff frequency of 0.02. (b) Cross-correlation of one half of the phased and filtered photometry with the other half. The red, blue, and green curves set the 92$\%$, the 95$\%$, and the 99$\%$ confidence limits, respectively, using 1000 repetitions.  A real signal should have
$>$1\% of the bins centered around zero lag above the 99\% confidence level; no such signal is present.}
\end{center}
\end{figure}

\begin{figure*}[ht]
\begin{center}
\includegraphics[scale=0.40, angle=0]{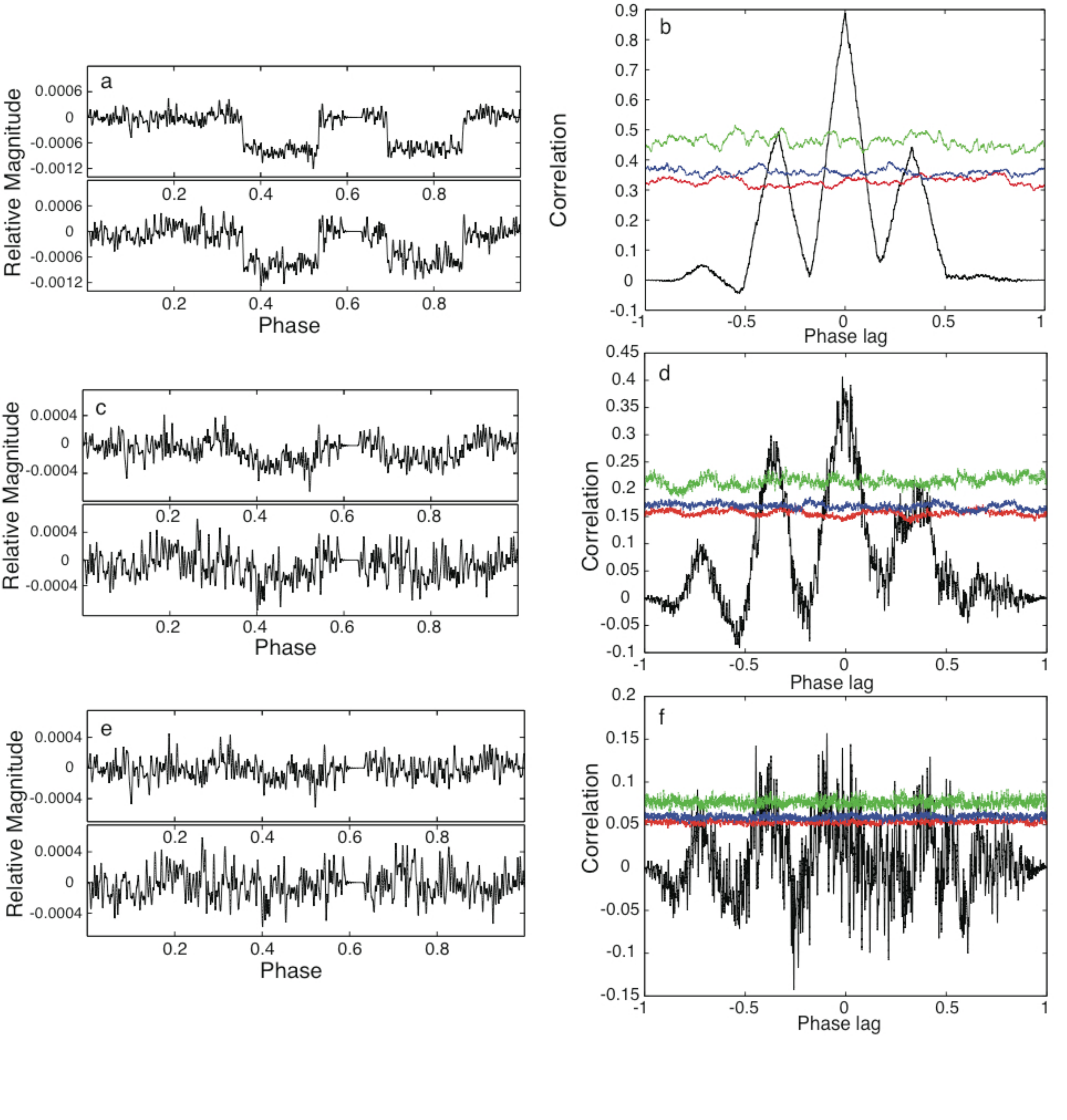}
\caption{Results from cross-correlation analysis of data with synthetic Trojan transits. The panels on the left-hand side show the two data halves (with artificial Trojan transits) that correspond to the cross-correlations shown in black on the right-hand side. Panels on the right also show the 92$\%$, 95$\%$, and 99$\%$ confidence intervals in red, blue, and green, respectively. In panel (a) the artificial Trojan transit depth is $8 \times 10^{-4}$, which shows up as a clear Trojan signal in the cross-correlation (panel b). Panel (c) has artificial transit depths of $2 \times 10^{-4}$. Its corresponding cross-correlation (panel d) shows peaks above the confidence intervals still clearly centered on the lags of -1/3 and 1/3. Panels (e) and (f) show the data and the cross-correlations, respectively, for a Trojan transit depth of $5\times10^{-5}$. In panel (f), the Trojan transits are not clearly discernible.}
\end{center}
\end{figure*}

\begin{figure}[htpb]
\begin{center}
\includegraphics[scale=0.55, angle=0]{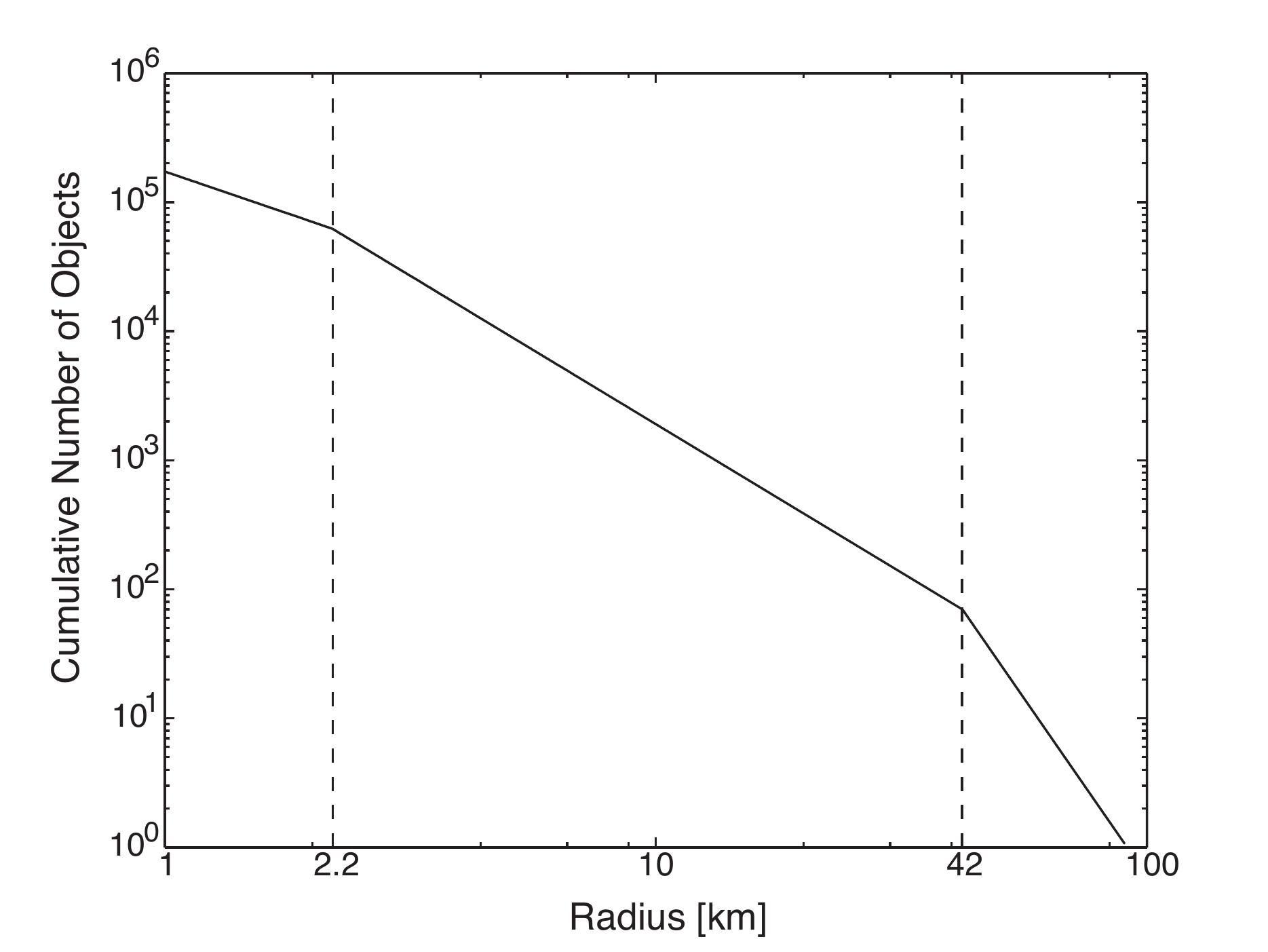}
\caption{Cumulative size distribution for solar system asteroids based on Jewitt et al. (2000) and Yoshida $\&$ Nakamura (2005). The breaks in the distribution occur near an asteroid 
radius of 2.2 km and 42 km. }
\end{center}
\end{figure}

\begin{figure}[ht]
\begin{center}
\includegraphics[scale=0.8, angle=0]{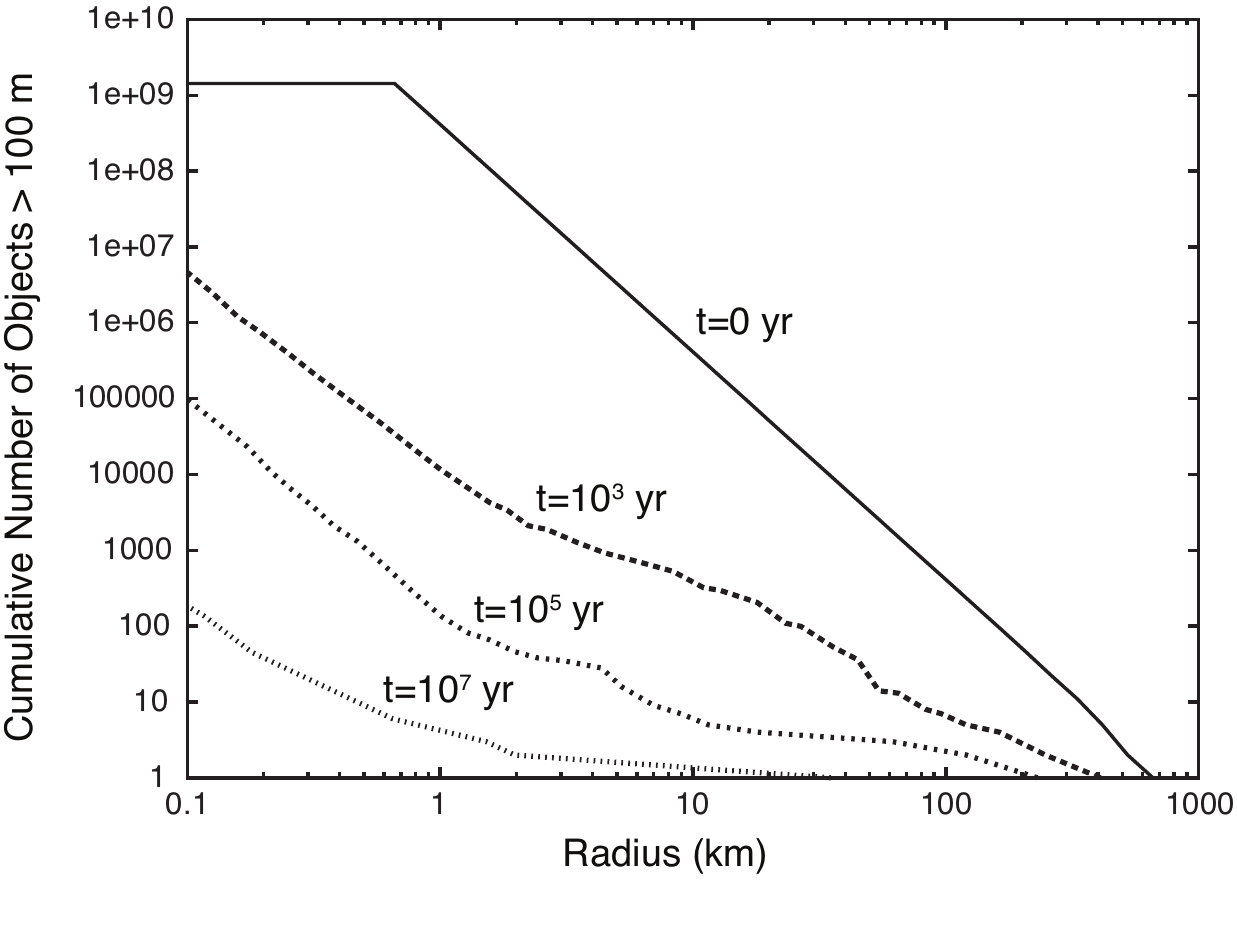}
\caption{ Results from dynamic collisional simulation for initial 1 lunar mass of Trojans. Here, the cumulative number of objects greater than 100 m is plotted as a function of object radius for various time steps in the simulation. By 10 Myr there is almost nothing left of the original size distribution (the program does not keep track of objects less than $\sim$ 100 m). Thus, from this plot we can see that the cloud grinds itself down to meter-sized objects within a few million years. }
\end{center}
\end{figure}

\begin{figure}[ht]
\begin{center}
\includegraphics[scale=0.8, angle=0]{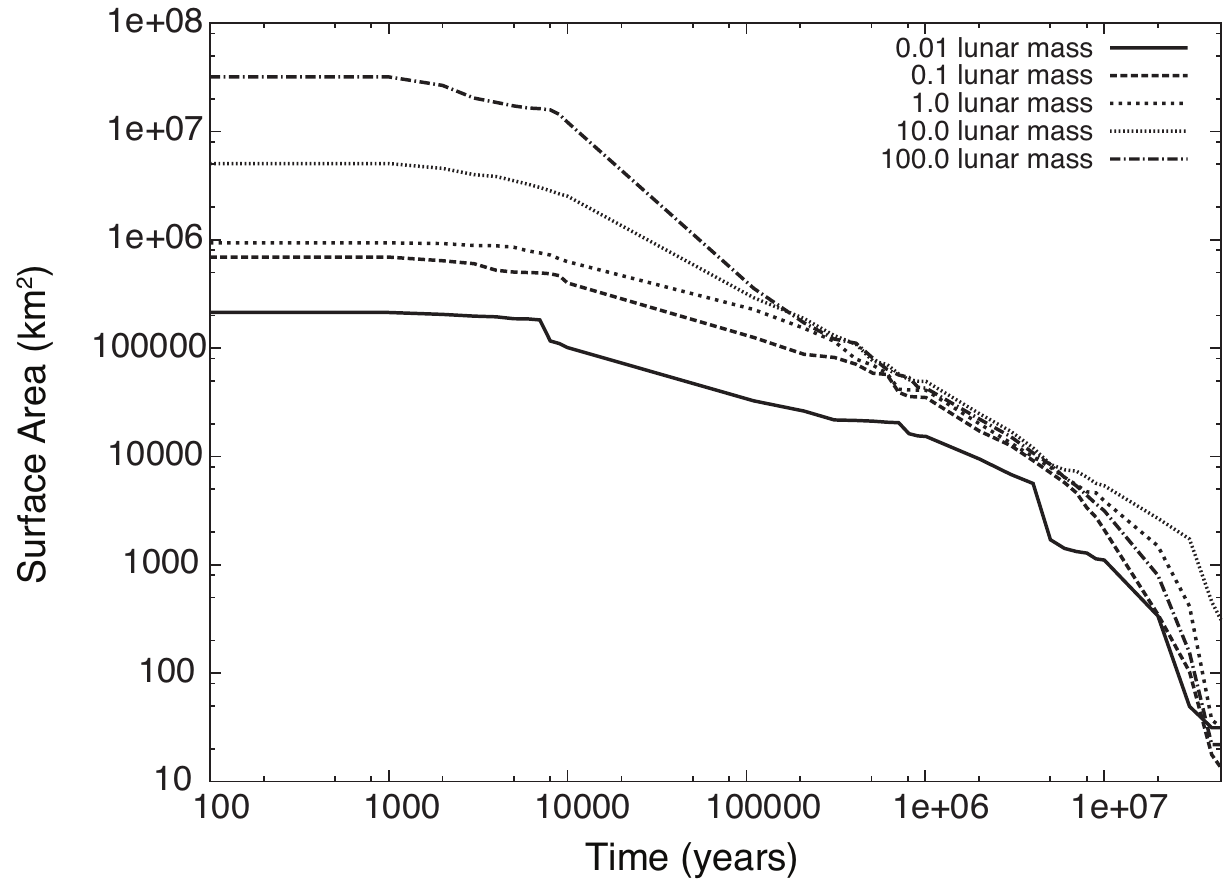}
\caption{Surface area of Trojan cloud as a function of time from simulation for various initial cloud masses. The initial total mass of the cloud (in lunar masses) is indicated beside each distribution. The total surface area (from objects with radii $>$50~m) of the cloud decreases with time independent of the initial starting mass. After $\sim$ 1 Myr, the surface areas for the different initial masses are the same within a factor of 3 of each other. }
\end{center}
\end{figure}

\end{document}